\documentstyle[psfig]{l-aa}
\begin{document}
\thesaurus{01(11.06.2; 11.19.7; 12.04.1; 12.07.1)}
\title{Dark Galaxies -- the Dominant Population?}
\author{M.R.S. Hawkins}
\institute{Royal Observatory,
Blackford Hill,
Edinburgh EH9 3HJ,
Scotland, UK}

\date{Received July 25; accepted October 2, 1997}

\maketitle

\begin{abstract}

This paper argues that there is a large population of dark galaxies 
which reveals its presence by the gravitational lensing of quasars,
and outnumbers normal galaxies by around 3:1.  There are 8 double
quasars with a separation greater than 2 arcseconds which have been
classified as probable or certain gravitational lenses.  Lensing
galaxies have only been found for 2 of these systems, and analysis of
the remainder has led in each case to the conclusion that they are
best explained by the gravitational lensing effect of a `dark galaxy'.
Here we examine the ensemble properties of this sample and conclude
that there is overwhelming evidence that the systems are indeed being
gravitationally lensed by dark galaxies or perhaps dark matter
galactic halos.  The existence and nature of such objects raises some
intriguing questions, as well as having profound implications for
large scale structure.  Several of the quasar systems show strong
evidence for microlensing.  It is argued that this implies a
substantial component of dark matter in the form of compact bodies,
either in the halo of the lensing galaxy or more generally along the
line of sight.


\end{abstract}

\section{Introduction}

In this paper we examine the small but well-known class of double
quasars where a quasar image has apparently been split into two by
the gravitational lensing effect of a galaxy mass object along the
line of sight.  In most cases the
lensing galaxy is not detectable, and the tight limits on the
mass-to-light ratios suggest that the galaxies may be `dark' in the
sense that they have failed to form stars.  This could be because they
do not contain baryonic material, or more plausibly because the
conditions for extensive star formation are not present.

In order to establish the existence of the dark galaxies, it is
necessary to make the case that the double quasars are
lensed and not binary systems, and that the lensing galaxy is truly
dark and not concealed in some way.  For any particular quasar pair it is
sometimes possible to argue special circumstances which circumvent the
need to invoke a dark galaxy.  In this paper we use the statistical
properties of the known systems to argue that there is indeed
a population of dark galaxies to be explained.

\section{The quasar sample}

At present there are 8 double quasars known with a separation greater
than 2 arcseconds that are plausible lens systems (Walsh et al. 1979;
Surdej et al. 1987; Weedman et al. 1982; Djorgovski \& Spinrad 1984;
Meylan \& Djorgovski 1989; Hewett et al. 1989; Wisotski et al. 1993;
Hawkins et al. 1997).  The properties of these systems are summarised
in Table 1, and include the redshift, the separation in arcseconds,
the magnitudes of the two components in the $R$ band, and the velocity
difference.  The last 3 columns refer to the lens and are discussed
later.  It will be seen that all systems have an
\begin{center}
\begin{table*}
\caption[]{Parameters for double quasar systems}
\begin{tabular}{ccccccrrlcr}
Name&z&sep&m$_A$&m$_B$&$\delta$m&\multicolumn{2}{c}{$\delta$v}&R$_g$&
M$_g$&M/L\\
&&($''$)&&&&\multicolumn{2}{c}{(km sec$^{-1}$)}&&$10^{12}_{\odot}$&\\
&&&&&&&&&\\
 Q0957$+$561& 1.41 & 6.1 & 16.6 & 17.0 & 0.4 &   3 & $ \pm$14&18.5&1.8&
   26\\
 Q0142$-$100& 2.72 & 2.2 & 16.9 & 19.1 & 2.2 & -24 & $\pm$109&19  &0.2&
    4\\
&&&&&&&&&\\
 Q2345$+$007& 2.15 & 7.0 & 18.9 & 20.4 & 1.5 &  15 & $ \pm$20&26  &2.6&
22000\\
 Q1635$+$267& 1.96 & 3.8 & 19.1 & 20.7 & 1.6 &  41 & $ \pm$54&23.5&0.7&
  590\\
 Q1120$+$019& 1.46 & 6.5 & 16.5 & 21.2 & 4.7 & 200 & $\pm$100&23  &2.0&
 1100\\
 Q1429$-$008& 2.08 & 5.1 & 17.7 & 20.8 & 3.1 & 260 & $\pm$300&24  &1.3&
 1700\\
 Q1104$-$180& 2.30 & 3.0 & 16.7 & 18.6 & 1.9 &   0 & $ \pm$90&24  &0.4&
  540\\
 Q2138$-$431& 1.64 & 4.5 & 19.8 & 21.0 & 1.2 &   0 & $\pm$115&23.8&1.0&
 1100\\
\end{tabular}
\end{table*}
\end{center}

image separation less than 8 arcseconds.  The surface density of
quasars to a magnitude
limit of $B = 21$ is about 30 per square degree (Hawkins \& V\'{e}ron
1995).  This implies that the probability of a given quasar having a
companion at a distance of $2 - 7$ arcseconds is in the range
$10^{-4} - 10^{-5}$.  Thus in a typical search of 1000 candidates there
is a probability of 1\% to 10\% of finding a close pair by chance, not
a particularly unlikely outcome.  This will be made more likely by the
effects of clustering, and less likely by the additional requirement
for the redshifts to be the same in a lensed system.  Various selection
effects will further change the probability, but it seems unlikely that
random associations can be convincingly ruled out on statistical
grounds in any particular case (see for example Hawkins et al. 1997).

\section{Tests for gravitational lensing}

The existence of the small sample of lens candidates in Table 1 makes it
possible to carry out a different test.  The distribution of separations
is shown in Fig. 1(a).  If the double quasars are chance associations,
one would expect the histogram of separations to increase linearly.
Selection effects will tend to increase this trend, as close pairs are
typically the hardest to find.  In fact the distribution falls to zero
beyond 8 arcsecs, even though most surveys are aimed at detecting
associations to at least 20 arcsecs (Webster et al. 1988;
Reimers 1990; Sramek \& Weedman 1978).  At greater distances the
probability of chance coincidences is no longer negligible, although as
Schneider (1993) points out it is perhaps surprising that even now very
few other quasars with similar redshifts are known with separations less
than two arcminutes.  If there are indeed no systematic effects in the
selection, then the hypothesis that the observed separations are
consistent with chance associations can be ruled out as completely
negligible by a Kolmogorov-Smirnoff test.  If one adopts a correlation
function (Collins et al. 1992) of the form $r^{-0.7}$ this
flattens the expected relation to $r^{0.3}$ but this is still
inconsistent with the data at a very high confidence level.
\begin{figure}
\hspace*{0.5cm}
\psfig{figure=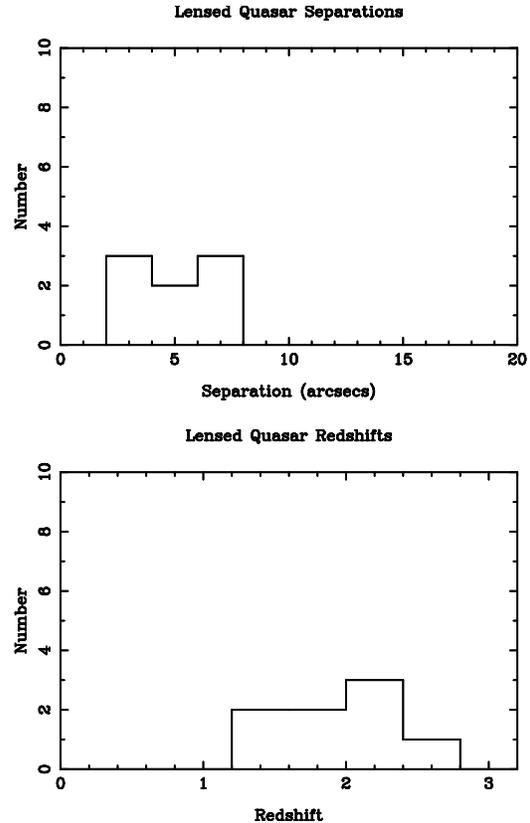,height=11cm,angle=0}
\caption[]{Histograms of the properties of the double quasar systems
in Table 1.  The top panel shows the distribution of separations of
the two components and the bottom shows the distribution of
redshifts.}  
\end{figure}
If on the other hand quasars are gravitationally lensed, the
separation is basically determined by the mass of the lens.  To
produce the largest separation of about 7 arcseconds a lensing mass of
$3 \times 10^{12}M_{\odot}$ is required, similar to that of the most
massive galaxies known.  This statistical argument is only modified
for binary quasars formed in the same protogalactic environment
if there is some sort of mutual triggering mechanism for quasar
activity, an idea which is at present of a speculative nature.

Fig. 1(b) shows a histogram of redshifts of the quasars in Table 1.
It will be seen that the distribution is highly asymmetric.  All 8
quasars have a redshift $z > 1.4$ in contrast to the expected
distribution which is almost flat (Hewett et al. 1993).  Although
a proper statistical test is difficult to carry out as several
different search procedures were used (Webster et al. 1988;
Reimers 1990; Hawkins et al. 1997; Sramek \& Weedman 1978)
with possibly different redshift biases, none of them would
appear to discriminate significantly against low redshift objects, and
there is no lack of single low redshift quasars.  On the other hand,
the observed redshift distribution agrees well with that expected on the
basis of gravitational lensing probabilities (Turner 1990), which drop
sharply below a redshift $z = 1$. The
velocity difference between the two components in all the systems in
Table 1 is consistent with zero, and in all but two cases has an upper
limit of about 100 km sec$^{-1}$.  The observed pairwise velocity
dispersion is (Ratcliffe et al. 1996) about 400 km sec$^{-1}$.
This is much larger than the observed limits, but must be reduced to
allow for evolutionary effects.  This test can be improved considerably
when tighter limits can be placed on the velocity differences.

The claims that systems in the bottom part of Table 1 are lensed rest
largely on detailed comparisons of the spectra of the two components
(Meylan \& Djorgovski 1989; Hewett et al. 1989; Turner et al. 1988;
Steidel \& Sargent 1991; Wisotski et al. 1995).  It is generally
accepted that small differences in the continuum
and absorption line systems can be accommodated within the lensing
picture as effects of time delay, microlensing and different light
paths for the two components.  Nonetheless, spectra of the two
components are found to be very similar, and the emission line systems
all but identical.  This raises the question of how similar one might
expect two arbitrary quasars at the same redshift to be.  Although it is
still difficult to answer this question properly (Turner et al. 1988),
one can quite easily compare the colours of the two components
with a sample of quasars at the same redshift.  There are two systems
in Table 1 which have accurate multicolour CCD photometry, and these
are plotted in Fig. 2 together with all quasars with a similar redshift
from the survey of Hawkins \& V\'{e}ron (1995).
Although the two components of the double quasars have identical
colours within the measurement errors, there is a wide spread in colour
for other quasars at the same redshift.

\section{Lensing by dark galaxies?}

Taking together the individual analysis of each double quasar with the
ensemble properties discussed here, there appears to be an overwhelming
case that the systems in Table 1 are gravitationally lensed.  This then
raises the question of the nature of the lensing objects.  For the
first two systems, lensing galaxies are clearly visible at $z = 0.39$
and $z = 0.45$, near the most probable value (Turner et al. 1984).
The lensing object lies close to the fainter component, as
expected for lensed systems.  For the remainder, in spite of intense
efforts (eg Tyson et al. 1986) no lensing galaxies have been
found.  Limits of $R > 23$ to $R > 26$ have been put on the magnitude
of any possible lens, implying mass-to-light ratios in excess of
500 $M_{\odot}/L_{\odot}$.  The last 3 columns in Table 1 show the
R-band magnitude limit placed on a possible lensing galaxy, the mass of
the lens and the resulting mass-to-light ratio, assuming 
$H_{0} = 50$km sec$^{-1}$Mpc$^{-1}$.  The mass of the lens is based on
the separation of the two components, and where a lensing galaxy is
undetected assumes a lens at $z = 0.5$, the most probable redshift
(Turner et al. 1984).  The M/L can only be reduced significantly
by putting the lens close to the quasar.  This is a very unlikely
configuration (Turner et al. 1984), and requires a large
increase in the mass of the lens.  The mass-to-light ratios in Table 1
are at least 10 to 100 times larger than for normal galaxies
(White 1990), which can effectively be ruled out as lens candidates.

  It has been known for some time (Narayan et al. 1984) that
diffuse mass distributions such as galaxy clusters can in principle
produce multiple images of quasars.  This can be done either by the
cluster on its own, or in combination with a galaxy.  In the first case
fine tuning is required to produce separations of a few arcseconds,
but this may be partly offset by the effects of amplification bias.
One would also expect the closer quasar pairs to be brighter, a trend
which is not evident in the systems in Table 1.  The presence of a
galaxy between the two images combined with an increase in surface
mass density from a cluster can produce a larger separation than
would be seen from the galaxy alone, suggesting a larger
mass-to-light ratio.  This picture has been suggested as an
explanation for the wide separation system Q2345+007 where shear
has been detected and a candidate cluster is visible
(Pell\'{o} et al. 1996).
There should however be a third fainter quasar image between
the two brighter images which has not so far been detected.

\begin{figure*}
\psfig{figure=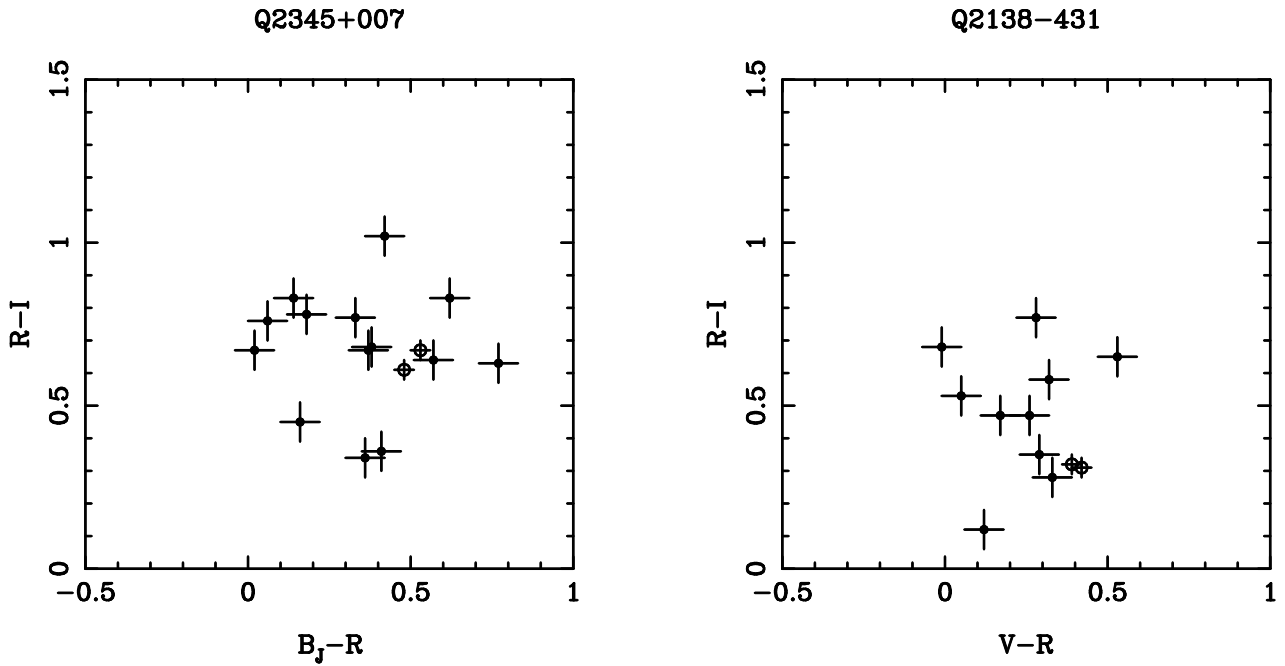,height=14cm,angle=270}
\vspace*{-2.5cm}
\caption[]{Two colour diagrams for quasars.  The open circles are CCD
measures of the two components of double systems, and the filled
circles are photographic measures of other quasars with similar
redshifts ($\delta z < 0.06$).}
\end{figure*}

It is hard to escape the conclusion that `dark galaxies', or perhaps
dark matter galactic halos, are responsible for lensing 6 out
of the 8 quasars in Table 1.  If we accept that the quasar lenses
represent a fair sample of galaxies we must conclude that 3 in 4
galaxies are dark, in the sense that they have a mass-to-light ratio
of at least several hundred $M_{\odot}/L_{\odot}$.  A mechanism for
formation and evolution of such galaxies has recently been described
by Jimenez et al. (1997).

  In most of the systems there is evidence for microlensing
(Wisotski et al. 1993; Hawkins et al. 1996;
Steidel \& Sargent 1991; Schild 1996).  This could be caused by stars
or other compact objects in the lensing galaxy, but would require an
optical depth to lensing approaching unity to produce a high
probability of variation (Kayser et al. 1986; Schneider \& Weiss 1987).
Thus the total dark matter content of the galaxy would have to
be in the form of microlensing bodies (Kayser et al. 1986), and
the stellar population would not be sufficient except perhaps close to
the nucleus.  Even if the lensing galaxy were entirely composed of
microlensing bodies it typically lies close to the fainter of the two
quasar images, and is not in a position to lens the brighter image.
It is perhaps more plausible that the microlensing arises from a
general distribution of dark matter bodies along the line of sight
(Hawkins 1996), but either way it supports the idea of dark matter in
the form of compact bodies.

\section{Conclusions}

  In this paper we have defined a sample of double quasars which are
plausible gravitational lens candidates.  In each individual case
earlier papers have made strong but not conclusive arguments that they
are indeed gravitationally lensed systems.  Here we consider the
ensemble properties of these candidates from a statistical point of
view and conclude that there is an overwhelming case that most if not
all the quasars are lensed.  6 out of 8 of the systems contain no
detectable lensing galaxy, implying a minimum mass-to-light ratio
of several hundred, and suggesting that dark galaxies may outnumber
normal ones by a substantial amount.

  Most of the double quasar systems show evidence for microlensing.
If this is caused by compact bodies in the lensing galaxy it would
imply that the dark halo was made up almost entirely of substeller
compact bodies.  A more plausible picture may be one in which the
microlensing is taking place all along the line of sight, and double
systems are no different from normal quasars in this respect.

\end{document}